\documentclass{ceab}   

\usepackage{epsfig}     
\usepackage{graphicx}   
\usepackage{lipsum}

\usepackage{ceabbib}     
\usepackage[T1]{fontenc}

\begin{document}

\def\tit{The strange close binary V405 And}
\def\aut{K. Vida et al.}

\title{
The strange close binary V405 And: the influence of different dynamos on the components
}

\author{
K. Vida$^1$,
L. Kriskovics$^1$,
K. Ol\'ah$^1$,
H. Korhonen$^2$
\vspace{2mm}\\
\it $^1$ Research Centre for Astronomy and Earth Sciences, \\
\it Hungarian Academy of Sciences\\
\it $^2$ Niels Bohr Institute, University of Copenhagen, \\
\it Juliane Maries Vej 30, 2100 K\o benhavn \O, Denmark
}

\maketitle

\begin{abstract}
V405 And is a fast-rotating ($P=0.465$d) grazing eclipsing binary with two active components. Using multicolor photometric monitoring and radial velocity measurements, we find that
the primary and the secondary components have masses of 0.49 and 0.21 solar masses, meaning that the primary probably possesses a radiative core and a convective envelope, while the secondary is fully convective. The radius of the low-mass component fits well the theoretical mass-radius relation. However, the radius of the primary is significantly larger than the predicted value, in fact, the discrepancy is far the highest from the four similar objects known. 
\end{abstract}

\keywords{Hvar astrophysical colloquium - proceedings - instructions}

\section{Introduction}

V405~And is an active binary system, which was first studied by \cite{chil}, who analyzed spectroscopy and photometric observations in $B$ and $V$ passbands, and found a rotation period of 0.465 days, a grazing eclipse, and strong H$\alpha$ emission. They also concluded that -- according to the H$\alpha$ region -- both components are active. 

 \cite{vida09} studied photometry in $BV(RI)_C$ passbands, and spectroscopic measurements. The authors found that the primary component has radius that is too large for its mass, while the secondary component fits quite well to the mass--radius relation.
 
 Their results were later confirmed by \cite{ribeiro}, who found very similar radius values using an independent spectroscopic method based on Doppler imaging.

\section{Results from binary modeling}
\begin{figure}
\centering
\includegraphics[angle=-0,width=0.45\textwidth]{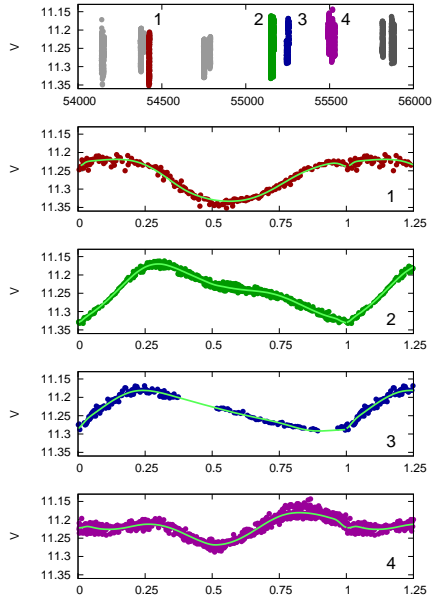}
\caption{$V$ light curve of V405~And (top), and phased $V$ light curves from different observing runs, together with the binary+spot model fit.}
\label{fig:lc}
\end{figure}

\begin{figure}
\centering
\includegraphics[angle=-90,width=0.70\textwidth]{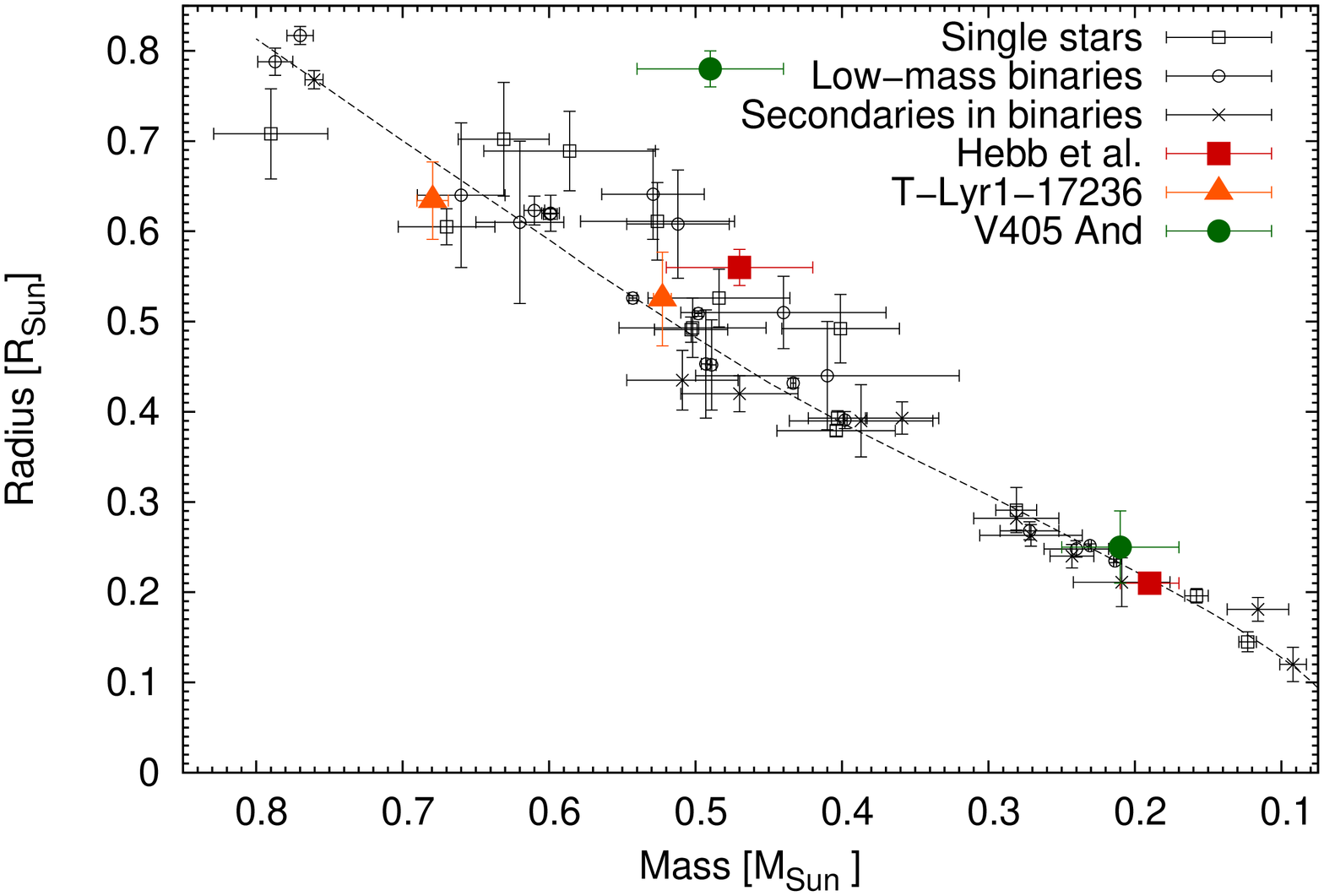}
\caption{Mass--radius curve from \cite{1998AA...337..403B} and observed values from  
\cite{data1,data2,data3,data4}  . Three systems, a binary from NGC 1647 \citep{hebb}, T-Lyr1-17236 \citep{tlyr}, and V405 And (present paper) are plotted with special characters.
}
\label{fig:MR}
\end{figure}


Figure \ref{fig:lc} shows photometric measurements in $V$ passband covering about 1800 days. We used the same iterative modeling method described in \cite{vida09} to fit the new observations. The basic idea is to create an initial binary model using PHOEBE \citep{phoebe} based on the radial velocity data of \cite{chil} and the light curve, which rudimentarily describes the effects of binarity. This model is subtracted from the observed light curve, resulting in a dataset, where the changes are caused by the spottedness. This data is then modeled by {\sc SpotModeL} \citep{sml}. By subtracting this model from the observed light curve we get a dataset, where the changes are due to the binarity (eclipse and proximity effects), which is fed again to PHOEBE. The previous steps are then repeated, until the combined binary and spot model describes the observed light curve satisfactorily. 
The bottom four panels of Figure \ref{fig:lc} show different parts of the observations together with the combined binary and spot model. 

We found that all the previous and newly observed light curves can be described by the same binary model that was found in \cite{vida09} confirming the results of that paper, namely, that the radius of the primary component is much larger than expected, while the secondary fits well to the mass--radius relation (see Fig. \ref{fig:MR}). There are more known systems that show similarly a larger radius for the primary component, but the discrepancy is the largest in the case of V405~And. 

We do not know, what is behind this behavior. \cite{activerad} showed that very strong magnetic field (which manifests itself in activity features) increases stellar radii, although not to this extent.  
\cite{morales} showed that starspots in certain cases can cause systematic deviations of a few percent in the determined stellar radii during modeling, but the difference in the case of V405~And is higher than that, too. 
The components have different structures: the secondary has a mass of $0.21M_\mathrm{Sun}$, thus it is probably fully convective. The primary has a mass of $0.49M_\mathrm{Sun}$, so it has a radiative core and a convective envelope. If our results and the similar values of \cite{ribeiro} are correct, the study and modeling of similar active systems, with components of different structures could be very interesting in the future.

\section{Long-term behavior}

\begin{figure}
\centering
\begin{minipage}[h]{0.47\linewidth}
\centering
\includegraphics[angle=-90,width=\textwidth]{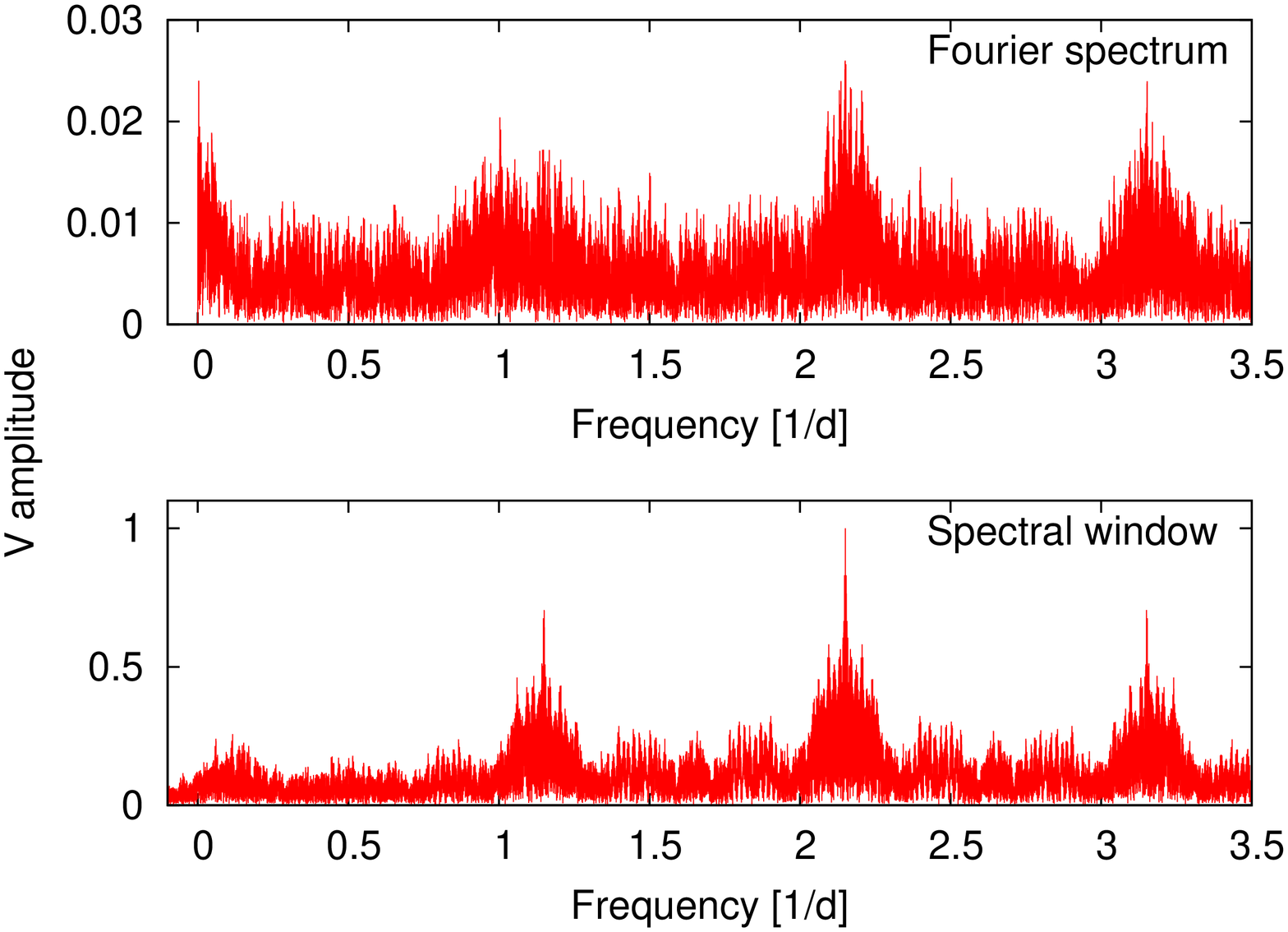}
\end{minipage}
\begin{minipage}[h]{0.47\linewidth}
\centering
\includegraphics[angle=0,   width=\textwidth]{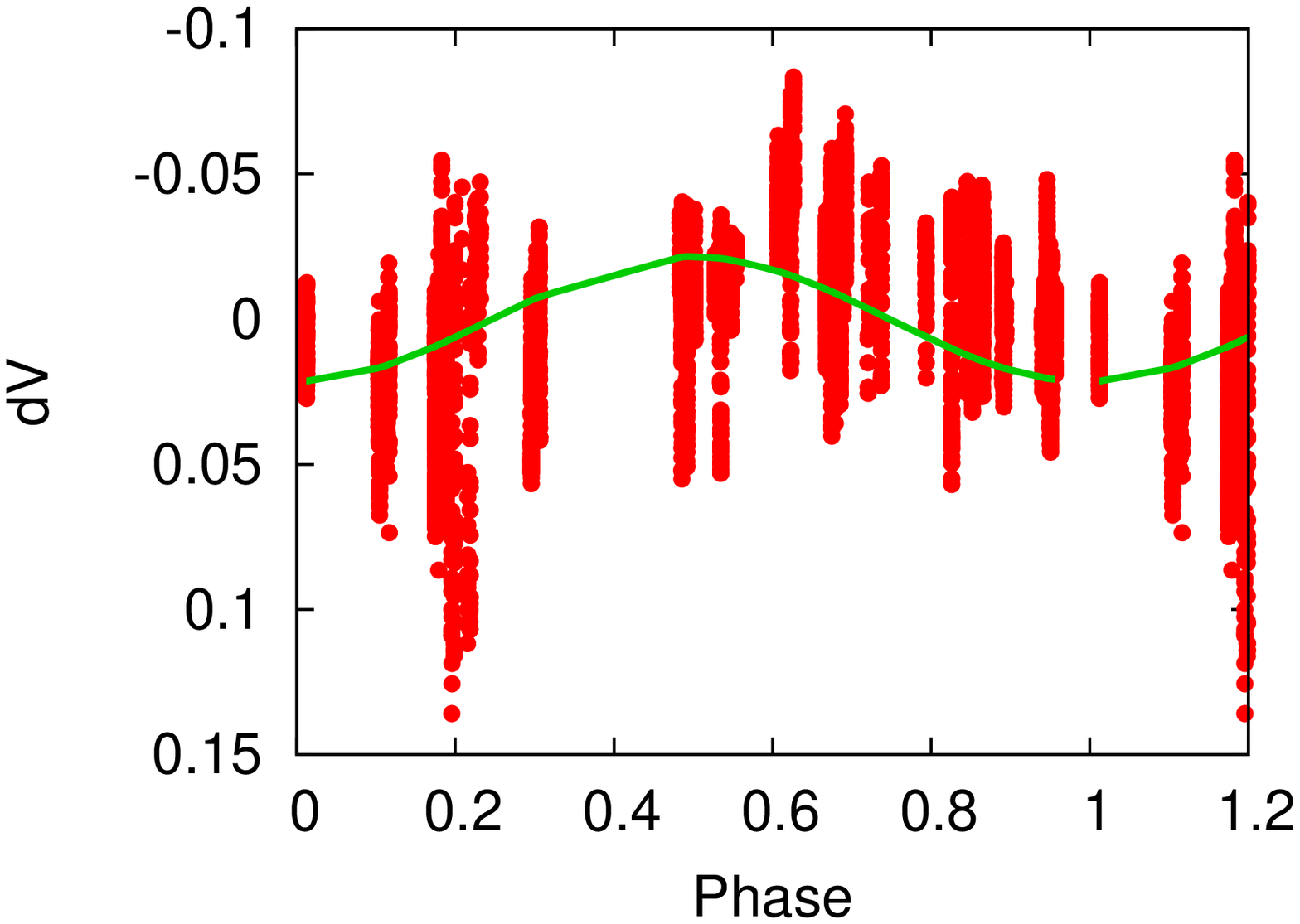}
\end{minipage}
\caption{Left: Fourier spectrum of the $V$ light curve showing the signal of the rotational modulation/orbital period at 0.465 days, and a peak at about 300 days, a result of an activity cycle (top), and the spectral window (bottom). Right: $V$ light curve pre-whitened with the rotational modulation, phased with the period of the activity cycle. }
\label{fig:cyc}
\end{figure}

We studied the $V$ light curve by discrete Fourier analysis using MUFRAN \cite{mufran}. The resulting Fourier spectrum and spectral window is plotted in Fig. \ref{fig:cyc}). The spectrum shows two main features. One can be associated with the rotational/orbital modulation at the frequency corresponding to the period of 0.465 days. The other one is a longer cycle, with a period of about 300 days. This is most probably a result of a stellar activity cycle, which is similar to the 11 year-long cycle of the Sun. As a result of the ultrafast rotation, the length of the stellar activity cycle is also very short (see \citealt{2002AN....323..361O}). The cycle found on V405~And is one of the shortest cycles ever found (cf. \citealt{eydra,shortcyc}).

\section*{Acknowledgements} 

The financial support of the OTKA grant K-81421, is acknowledged.
This work was also supported by the ``Lend\"ulet-2009'', and ``Lend\"ulet-2012'' Young Researchers' Programs of the Hungarian Academy of Sciences, and by the HUMAN MB08C 81013 grant of the MAG Zrt.
We would like to thank the anonymous referee for his helpful comments.

\bibliographystyle{ceab}

\end{document}